\documentclass{eptcs}
\usepackage[utf8]{inputenc}
\usepackage[british]{babel} 

\usepackage{etoolbox} 
\newtoggle{techreport}
\togglefalse{techreport}

\usepackage{afterpage}              
\usepackage{amsmath}
\usepackage{amssymb}
\usepackage{amsthm}
\usepackage{appendix} 
\usepackage{cancel}
\usepackage{caption}                
\usepackage{cmll} 
\usepackage{colonequals}            
\usepackage{DotArrow} 

\usepackage{enumitem}        

\usepackage{environ}                
\usepackage{epigraph} 
\usepackage{etex,etoolbox}          
\usepackage{eucal} 

\usepackage[nomargin,inline,index]{fixme} 
\fxusetheme{color}
\FXRegisterAuthor{fxAS}{anfxAS}{\color{blue} {\underline{Alceste}}}
\FXRegisterAuthor{bart}{anbart}{\color{magenta} {\underline{bart}}}

\usepackage{float} 
\usepackage{framed} 
\usepackage{graphicx}               

\usepackage[protrusion=true,expansion=true]{microtype} 

\usepackage{lipsum} 

\usepackage[%
  usenames,%
  dvipsnames%
]{xcolor} 

\usepackage{float}

\definecolor{listingBG}{HTML}{FFFFCB}%
\definecolor{listingFrame}{HTML}{BBBB98}%
\newsavebox{\mybgbox}%
\usepackage{multirow}               

\usepackage{mathtools} 
\usepackage{nicefrac}
\usepackage{pifont} 
\usepackage{proof}                  
\usepackage{relsize} 
\usepackage{rotating}               
\usepackage{stmaryrd}               
\usepackage{caption,subcaption}     
\usepackage{keyval}

\usepackage{thmtools, thm-restate, thm-autoref}  
\usepackage{trsym} 

\usepackage{fixltx2e} 
\usepackage{xspace} 
\usepackage{xifthen} 

\usepackage{wrapfig} 

\usepackage{hyperref} 
\hypersetup{final} 
\usepackage{cleveref}

\newif\ifdraft%
  \draftfalse%
  \drafttrue%

\newcommand{\ifempty}[3]{%
  \ifthenelse{\isempty{#1}}{#2}{#3}%
}

\def\etc{\emph{etc.}\@\xspace}%
\def\eg{e.g.\@\xspace}%
\def\ie{i.e.\@\xspace}%

\def\RelEng{\texttt{LTSwb}\xspace}%

\def\enumroman{\emph{(\roman*)}}

\newcommand{\ltsLTS}[1][]{\mathord{\color{OliveGreen}\mathbb{L}_{#1}}}%
\newcommand{\ltsStates}{\Sigma}%
\newcommand{\ltsState}{\sigma}%
\newcommand{\ltsStatei}{\sigma'}%
\newcommand{\ltsLabs}[1][]{\mathord{\color{Brown}\Lambda_{#1}}}%

\def\InTag{\mathord{?}}%
\def\OutTag{\mathord{!}}%

\newcommand{\atomFmt}[1]{\mathsf{#1}}%
\newcommand{\atom}[2][]{\atomFmt{#2}_{#1}}%

\newcommand{\ltsLab}[1][]{\mathord{\color{Brown}\ell_{#1}}}%
\newcommand{\ltsIn}[2][]{\mathord{{\InTag{\atom[#1]{#2}}}}}%
\newcommand{\ltsOut}[2][]{\mathord{{\OutTag{\atom[#1]{#2}}}}}%
\newcommand{\ltsTau}[1][]{\mathord{\tau}}%

\newcommand{\relR}{\mathrel{\mathcal{R}}}
\newcommand{\relRi}{\mathrel{\mathcal{R}_1}}

\newcommand{\relRii}{\mathrel{\mathcal{R}_2}}

\def\code#1{{{\color{violet}\texttt{#1}}\xspace}}%
\def\codeStr#1{{\color{Mahogany}\texttt{\symbol{34}{#1}\symbol{34}}}\xspace}%

\def\ltsBaseFmt#1{#1}
\newcommand{\ltsSp}[1][]{\mathord{\ltsBaseFmt{p}_{#1}}}
\newcommand{\ltsSpi}[1][]{\mathord{\ltsSp[#1]\!'}}
\newcommand{\ltsSpii}[1][]{\mathord{\ltsSp[#1]\!''}}

\newcommand{\ltsSq}[1][]{\mathord{\ltsBaseFmt{q}_{#1}}}
\newcommand{\ltsSqi}[1][]{\mathord{\ltsSq[#1]\!'}}

\newcommand{\ltsMove}[2][]{\ifempty{#1}{\xrightarrow{#2}}{\mathrel{\xrightarrow{#2}_{\lts{#1}}}}}
\newcommand{\ltsNotMoveP}[2][]{\mathord{\not\ltsMove[#1]{#2}}} 
\newcommand{\ltsWMove}[2][]{\ifempty{#1}{\xRightarrow{#2}}{\mathrel{\xRightarrow{#2}_{\lts{#1}}}}}

\newcommand{\setcomp}[2]{\mathord{\left\{{#1} \,\middle|\, {#2}\right\}}}

\newcommand{\inferrule}[1]{\textsc{\scriptsize (#1)}}
\newcommand{\inference}[3][]{\infer[\ifempty{#1}{}{\inferrule{#1}}]{#3}{#2}}

\newcommand{\isomorph}{\cong}%

\def\procColor{BrickRed}
\newcommand{\proc}[1]{{\color{\procColor}#1}}
\def\procP{\mathord{\proc{P}}}

\newcommand{\procQueue}[1]{\mathord{\proc{\left[{#1}\right]}}}

\newcommand{\mytitle}{The LTS WorkBench}

\title{\mytitle}

\author{%
  Alceste Scalas%
  \institute{Dipartimento di Matematica e Informatica\\
    Università di Cagliari, Italy\\[1.2mm]%
    Department of Computing\\
    Imperial College London, UK}%
  \email{alceste.scalas@imperial.ac.uk}%
  \and%
  Massimo Bartoletti%
  \institute{Dipartimento di Matematica e Informatica\\
    Università di Cagliari, Italy}%
  \email{bart@unica.it}%
}

\def\titlerunning{\mytitle}
\def\authorrunning{A.\@ Scalas, M.\@ Bartoletti}

\usepackage{fancyvrb}
\input{pygments.tex}%

\newfloat{listing}{t}{listing}[section]%
\floatname{listing}{Listing}%
\crefname{listing}{listing}{listings}%
\Crefname{listing}{Listing}{Listings}%

\numberwithin{figure}{section}%

\begin{document}

\maketitle

\begin{abstract}
  Labelled Transition Systems (LTSs) %
  are a fundamental semantic model %
  in many areas of informatics, %
  especially concurrency theory. %
  Yet, reasoning on LTSs and relations between their states can be
  difficult and elusive: %
  very simple process algebra terms %
  can give rise to a large (possibly infinite) number of intricate
  transitions and interactions. %
  To ease this kind of study, %
  we present \RelEng, %
  a flexible and extensible LTS toolbox: %
  this tutorial paper %
  discusses its design and functionalities. %
\end{abstract}

\section{Introduction}
\label{sec:intro}%

\RelEng (from \emph{``\texttt{LTS} \texttt{W}ork\texttt{B}ench''}) %
\cite{LTSWB} %
is a Labelled Transition System (LTS) toolbox, %
allowing to define LTSs and processes, %
manipulate them, %
and compute relations between their states. %
Its main features are: %
\begin{description}
\item[genericity.] %
  \RelEng does not require LTSs and processes to have specific state/label
  types. %
  This allows %
  to semantically reason on different process specifications: %
  for example, %
  it allows to study whether 
  a CCS process \cite{Milner89ccs} %
  is a semantic refinement of a session type \cite{Honda93DyadicTypes} %
  (as in \cite{BartolettiSZ14Concur}), %
  or whether it can correctly interact with a service %
  whose specification is given as a Communicating Finite-State Machine (CFSM) %
  \cite{BZ1983CFSM};
\item[reusability.] %
  \RelEng is built upon an underlying \emph{relational calculus}, %
  whose operators allow for relation filtering, sequencing, parallel
  composition, \etc. %
  Such operators are fully generic, and can be reused \eg to implement
  different process calculi without having to redefine similar operators
  each time;
\item[laziness.] %
  Very large, %
  and even infinite-state LTSs and processes are managed
  transparently: %
  states and transitions are only generated upon request. %
  This allows to mitigate state space explosion problems, %
  and to explore and filter out (finite) parts of infinite LTSs
  arising \eg with recursion, parallelism, unbounded communication
  buffers, %
  \etc
\end{description}

\RelEng is a Scala \cite{Scala} library. %
The choice of Scala is motivated by %
the desire of a functional programming language with an advanced type
system, %
and an access to the vast landscape of libraries %
available on the Java VM; %
moreover, Scala's \code{lazy} values allow for controlled lazy
evaluation in an eager language %
--- a mix which we found helpful for our implementation.
\RelEng can  be used directly on the interactive Scala console: %
unless otherwise noted, %
all the examples on this paper can be replicated therein %
via simple cut\&pasting.

\section{LTSs, processes and asynchrony}
\label{sec:lts-processes}%

An \emph{LTS} is a triple %
$\left(\ltsStates, \ltsLabs, \mathord{\relR}\right)$ %
where $\ltsStates$ is the set of \emph{states}, %
$\ltsLabs$ is the set of \emph{labels}, %
and %
$\mathord{\relR} \subseteq%
\left(\ltsStates \times \left(\ltsLabs \times
    \ltsStates\right)\right)$ %
is the \emph{transition relation}. %
A \emph{process} is a pair $(\ltsLTS, \ltsState$) %
where $\ltsLTS$ is an LTS %
and $\ltsState$ is one of its states. %
The \emph{process transition} %
$(\ltsLTS, \ltsState) \ltsMove{\ltsLab} (\ltsLTS, \ltsStatei)$ %
holds iff %
$(\ltsState, (\ltsLab, \ltsStatei))$ %
is in the transition relation of $\ltsLTS$. %

In \crefrange{sec:from-lts-to-process}{sec:new-calculi} %
we show how %
\RelEng processes can be created %
(by extracting them from 
some LTS) %
and manipulated, %
and how the framework can be extended. %
But first, in \Cref{sec:relational-calculus} we give some intuition about the
underlying relational calculus. %
Note that such a section is not strictly
necessary to follow the rest of this tutorial paper, %
and it is possible to directly jump to \Cref{sec:from-lts-to-process}.

\subsection{Under the hood: a relational calculus}
\label{sec:relational-calculus}

In this section we sketch (and give reasons for) the \emph{relational
  calculus} at the core of \RelEng, %
by showing the correspondence between \emph{relational sequencing} %
and the well-known \emph{sequencing operator} %
provided by several process calculi.
We first need to introduce some more notation:
\begin{itemize}
\item%
  the \emph{set of continuations of a process after transition $\ltsLab$} %
  is\; %
  \(%
  \big(\ltsLTS, \ltsSp\big)(\ltsLab)%
  \;=\; %
  \setcomp{%
    \big(\ltsLTS, \ltsSpi)
    \,}{\,%
    \big(\ltsLTS, \ltsSp\big)
    \ltsMove{\ltsLab}
    \big(\ltsLTS, \ltsSpi\big)
  }
  \);%
\item%
  given a relation $\mathord{\relR} \subseteq \Delta \times \Gamma$, %
  the \emph{image of $\delta$ under $\relR$} is the set\; %
  \(%
  \mathord{\relR}(\delta)%
  \;=\; \setcomp{\gamma\,}{\,(\delta,\gamma) \in \mathord{\relR}}
  \).%
\end{itemize}

Now, say that we want to study the behaviour of processes %
written in a calculus $C$, %
equipped with the usual \emph{sequential composition operator} %
\code{($\ltsSp$ seq $\ltsSq$)}, %
with LTS semantics inductively defined by the rules: %
\[
\inference{%
  \text{\code{$\ltsSp$}} \ltsMove{\ltsLab} \text{\code{$\ltsSpi$}}
}{%
  \text{\code{($\ltsSp$ seq $\ltsSq$)}} \ltsMove{\ltsLab}%
  \text{\code{($\ltsSpi$ seq $\ltsSq$)}}
}
\qquad
\inference{%
  \text{\code{$\ltsSp$}} \ltsNotMoveP{}
  &
  \text{\code{$\ltsSq$}} \ltsMove{\ltsLab} \text{\code{$\ltsSqi$}}
}{%
  \text{\code{($\ltsSp$ seq $\ltsSq$)}} \ltsMove{\ltsLab}%
  \text{\code{($\ltsSp$ seq $\ltsSqi$)}}
}
\quad\text{(where\;} %
\text{\code{$\ltsSp$}} \ltsNotMoveP{} \text{ iff }%
\not\exists \ltsLab,\text{\code{$\ltsSpi$}}  \mathbin{.} %
\text{\code{$\ltsSp$}} \ltsMove{\ltsLab} \text{\code{$\ltsSpi$}}%
\text{)}%
\]

We can observe that such a definition is independent %
from the syntax of \code{$\ltsSp$}, \code{$\ltsSq$} and $\ltsLab$, %
and is therefore adaptable to different process calculi. %
Such a definition is also meaningful if, for example, %
\code{$\ltsSp$},\code{$\ltsSq$} are state-transition-state traces %
extracted from an execution log: %
their sequential execution would be defined in the same way.
Can we implement such a composition upon a \emph{reusable
  syntax-independent} foundation?

One way to address such a question is to \emph{define sequencing at an
  underlying relational level}. %
Let\;
$\mathord{\relRi} \subseteq \big(\ltsStates_1 \times (\ltsLabs[1]
\times \ltsStates'_1)\big)$ %
\;and\; %
$\mathord{\relRii} \subseteq \big(\ltsStates_2 \times (\ltsLabs[2] \times
\ltsStates'_2)\big)$. %
The \emph{sequencing of $\relRi$ and $\relRii$} %
is the  relation %
\[
\mathord{\relRi}\mathop{;}\mathord{\relRii} %
\,\subseteq\, %
\Big((\ltsStates_1 \times \ltsStates_2) \times \big((\ltsLabs[1] \cup
\ltsLabs[2]) \times (\ltsStates'_1 \times \ltsStates'_2)\big)\Big) %
\]
inductively defined by the following rules %
(notice the similarity with the \code{seq} rules above): %

\[
\inference{%
  \big(\ltsSp, \left(\ltsLab, \ltsSpi\right)\big)%
  \in \mathord{\relRi}%
}{%
  \Big((\ltsSp,\ltsSq),\, \big(\ltsLab, (\ltsSpi,\ltsSq)\big)\Big) %
  \,\in\, \mathord{\relRi}\mathop{;}\mathord{\relRii}%
}
\qquad%
\inference{%
  \mathord{\relRi}(\ltsSp) = \emptyset%
  &
  \big(%
  \big(\ltsSq, (\ltsLab, \ltsSqi)\big) %
  \in \mathord{\relRii}
}{%
  \Big((\ltsSp,\ltsSq),\, \big(\ltsLab, (\ltsSp,\ltsSqi)\big)\Big) %
  \,\in\, \mathord{\relRi}\mathop{;}\mathord{\relRii}%
}
\]

We can equivalently define the sequencing of $\relRi$ and $\relRii$ by
defining the image of $(\ltsSp,\ltsSq)$ %
under $\mathord{\mathord{\relR}_1 \mathop{;} \mathord{\relR}_2}$:
\[
\Big(\mathord{\mathord{\relR}_1 \mathop{;} \mathord{\relR}_2}\Big)%
\big((\ltsSp,\ltsSq)\big)%
\;=\;%
\begin{cases}
  \setcomp{\big(\ltsLab, (\ltsSpi, \ltsSq)\big)}{%
    \big(\ltsLab, \ltsSpi\big) \in \mathord{\relRi}(\ltsSp)%
  }
  &%
  \text{if $\mathord{\relRi}(\ltsSp) \neq \emptyset$}%
  \\[2mm]%
  \setcomp{\big(\ltsLab, (\ltsSp, \ltsSqi)\big)}{%
    \big(\ltsLab, \ltsSqi\big) \in \mathord{\relRii}(\ltsSq)%
  }
  &%
  \text{otherwise}
\end{cases}
\]

We can now ``lift'' the  sequencing operation %
from the relational level to the LTS level. %
Let\: %
$\ltsLTS[1] = \big(%
\ltsStates_1, \ltsLabs[1], \relR_1%
\big)$ %
\;and\; %
$\ltsLTS[2] = \big(%
\ltsStates_2, \ltsLabs[2], \relR_2%
\big)$. %
The \emph{sequencing of $\ltsLTS[1]$ and $\ltsLTS[2]$} is:
\[
\ltsLTS[1] \mathop{;} \ltsLTS[2] \;\;=\;\;%
\Big(\;%
\ltsStates_1 \times \ltsStates_2,\;\;%
\ltsLabs[1] \cup \ltsLabs[2],\;\;%
\mathord{\relR}_1 \mathop{;} \mathord{\relR}_2%
\;\Big)
\]
Finally, we can further lift sequencing to processes. %
The \emph{sequencing of processes %
  $(\ltsLTS[1], \ltsSp)$ and $(\ltsLTS[2], \ltsSq)$} is:
\[
(\ltsLTS[1], \ltsSp) \,\mathop{;}\, (\ltsLTS[2], \ltsSq)%
\;\;=\;\;%
\Big(\ltsLTS[1] \mathop{;} \ltsLTS[2]\,,\, (\ltsSp, \ltsSq)\Big)
\]
and we can observe that %
the process $(\ltsLTS[1], \ltsSp) \,\mathop{;}\, (\ltsLTS[2], \ltsSq)$
performs the transitions of $\ltsSp$ in $\ltsLTS[1]$, %
followed by those of $\ltsSq$ in $\ltsLTS[2]$.

\def\relRC{\mathrel{\mathord{\relR}_C}}%
We can now return to our calculus $C$, %
with its sequential composition\, %
\code{($\ltsSp$ seq $\ltsSq$)}. %
Let %
$\ltsLTS[C] = \big(%
\ltsStates_C, \ltsLabs[C], \relRC%
\!\big)$ %
be the LTS inhabited by $C$'s processes. %
We want the transition diagram of \code{($\ltsSp$ seq $\ltsSq$)} in
$\ltsLTS[C]$ %
to be 
\emph{observationally indistinguishable} 
from that of the sequenced processes 
$(\ltsLTS[C], \text{\code{$\ltsSp$}}) \,\mathop{;}\, (\ltsLTS[C], \text{\code{$\ltsSq$}})$, %
\ie:
\[
\Big(\ltsLTS[C], \text{\code{($\ltsSp$ seq $\ltsSq$)}}\Big)%
\;\isomorph\;%
(\ltsLTS[C], \text{\code{$\ltsSp$}}) \,\mathop{;}\, (\ltsLTS[C], \text{\code{$\ltsSq$}})
\qquad%
\text{%
  \footnotesize%
  (where $\isomorph$ is transition diagram equality, up-to node renaming)
}
\]
In other words, we want %
the continuations of \code{($\ltsSp$ seq $\ltsSq$)} after transition
$\ltsLab$ %
to be isomorphic to the continuations of %
$(\ltsLTS[C], \text{\code{$\ltsSp$}}) \,\mathop{;}\, (\ltsLTS[C], \text{\code{$\ltsSq$}})$ %
after $\ltsLab$. %
We can obtain this by requiring:
\begin{eqnarray}
  \nonumber%
  \Big(\ltsLTS[C], \text{\code{($\ltsSp$ seq $\ltsSq$)}}\Big)(\ltsLab)%
  &=&%
  \setcomp{%
    \Big(\ltsLTS[C], \text{\code{($\ltsSpi$ seq $\ltsSqi$)}}\Big)%
  \,}{\,%
    (\ltsLTS[C], \text{\code{$\ltsSpi$}}) \,\mathop{;}\, (\ltsLTS[C], \text{\code{$\ltsSqi$}})%
    \in%
    \Big((\ltsLTS[C], \text{\code{$\ltsSp$}}) \,\mathop{;}\, (\ltsLTS[C], \text{\code{$\ltsSq$}})\Big)(\ltsLab)
  }\\%
  \nonumber%
  &=&%
  \setcomp{%
    \Big(\ltsLTS[C], \text{\code{($\ltsSpi$ seq $\ltsSqi$)}}\Big)%
  \,}{\,%
    (\text{\code{$\ltsSpi$}}, \text{\code{$\ltsSqi$}}) \in%
    \Big(\ltsLTS[C]  \mathop{;}  \ltsLTS[C]\,,%
    (\text{\code{$\ltsSp$}},\text{\code{$\ltsSq$}})\Big)(\ltsLab)
  }\\%
  \label{eq:rc-seq-rc:i}%
  &=&%
  \setcomp{%
    \Big(\ltsLTS[C], \text{\code{($\ltsSpi$ seq $\ltsSqi$)}}\Big)%
  \,}{\,%
    (\text{\code{$\ltsSpi$}}, \text{\code{$\ltsSqi$}})%
    \in%
    \Big(\mathord{\relRC} \mathop{;} \mathord{\relRC}\Big)\big((\text{\code{$\ltsSp$}}, \text{\code{$\ltsSq$}})\big)(\ltsLab)
  }%
\end{eqnarray}
Since (by definition) %
\(%
\big(\ltsLTS[C], \text{\code{($\ltsSp$ seq $\ltsSq$)}}\big)(\ltsLab)%
=%
\setcomp{%
    (\ltsLTS[C], \text{\code{$\ltsSpii$}})%
  \,}{\,%
    \text{\code{$\ltsSpii$}}%
    \in%
    \mathord{\relRC}\left(\text{\code{($\ltsSp$ seq $\ltsSq$)}}\right)(\ltsLab)
  }%
\), %
we must also have:
\begin{equation}
  \label{eq:rc-seq-rc:ii}%
  \Big(\ltsLTS[C], \text{\code{($\ltsSp$ seq $\ltsSq$)}}\Big)(\ltsLab)%
  \;\;=\;\;%
  \setcomp{%
    \big(\ltsLTS[C], \text{\code{($\ltsSpi$ seq $\ltsSqi$)}}\big)%
    \,}{\,%
    \text{\code{($\ltsSpi$ seq $\ltsSqi$)}}%
    \in%
    \mathord{\relRC}\left(\text{\code{($\ltsSp$ seq $\ltsSq$)}}\right)(\ltsLab)
  }%
\end{equation}
Therefore, from (\ref{eq:rc-seq-rc:i}) and (\ref{eq:rc-seq-rc:ii}) %
we have that the image of \code{($\ltsSp$ seq $\ltsSq$)} %
under the transition relation $\relRC$ %
must be isomorphic to %
the image of $(\text{\code{$\ltsSp$}}, \text{\code{$\ltsSq$}})$ %
under the sequenced relation $\mathord{\relRC} \mathop{;} \mathord{\relRC}$\,:
\[
\mathord{\relRC}(\text{\code{($\ltsSp$ seq $\ltsSq$)}})%
\;\;=\;\;%
\setcomp{
  \big(\ltsLab, \text{\code{($\ltsSpi$ seq $\ltsSqi$)}}\big)
  \;}{\;
  \big(\ltsLab, (\text{\code{$\ltsSpi$}}, \text{\code{$\ltsSqi$}})\big) \in%
  \Big(\mathord{\relRC} \mathop{;} \mathord{\relRC}\Big)\big((\text{\code{$\ltsSp$}}, \text{\code{$\ltsSq$}})\big)%
}
\]

This last equation tells us that the transitions of
\text{\code{($\ltsSp$ seq $\ltsSq$)}} in $\ltsLTS[C]$ %
can be inductively defined upon the transitions of
$(\text{\code{$\ltsSp$}},\code{\code{$\ltsSq$}})$ in
$\mathord{\ltsLTS[C]} \mathop{;} \mathord{\ltsLTS[C]}$ %
simply by providing such an isomorphism %
--- which is just a syntactic deconstruction/reconstruction of the former
into/from the latter. %
Hence, the syntax-independent relational sequencing operator can
be reused (with minimal syntax-dependent additions) %
to define the sequencing operator at the process calculus level.

This theoretical foundation is the heart of the implementation of
\RelEng: all the LTS-level and process-level operators described in
the rest of this tutorial %
(including the more complex ones, %
such as parallel composition, asynchronous transformation and filtering)
are implemented upon underlying
syntax-independent relational operators. %

\subsection{From LTSs to processes}
\label{sec:from-lts-to-process}

\begin{figure}[t!]
  \begin{minipage}{0.5\linewidth}
    \centering
    \scalebox{0.8}{%
      \begin{minipage}{1.1\linewidth}
        \includegraphics[clip=true,trim=14mm 0 0 0,width=1.135\linewidth]{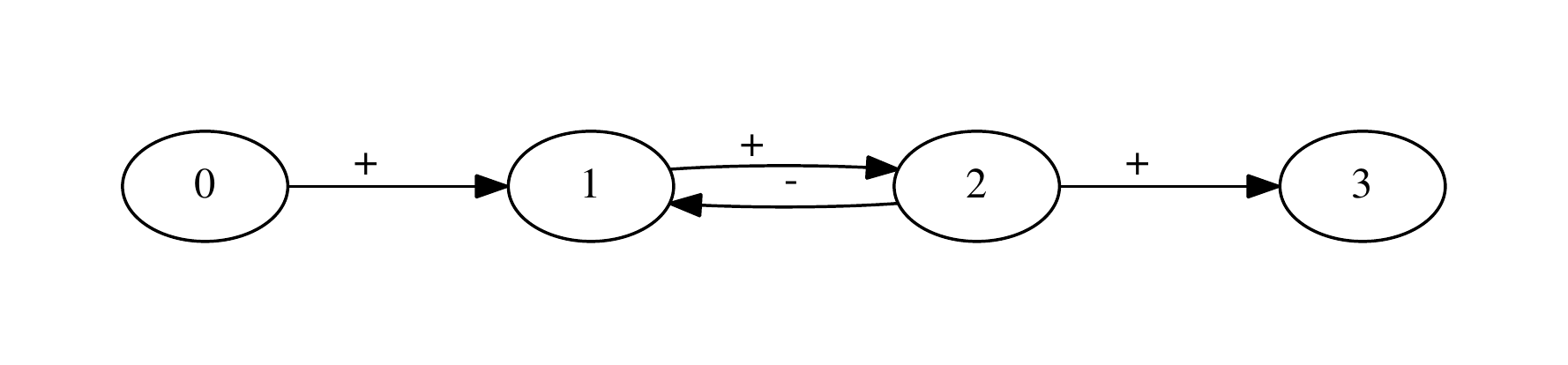}%
        \\%
        \vspace{-1.3cm}%
        \includegraphics[width=1.1\linewidth]{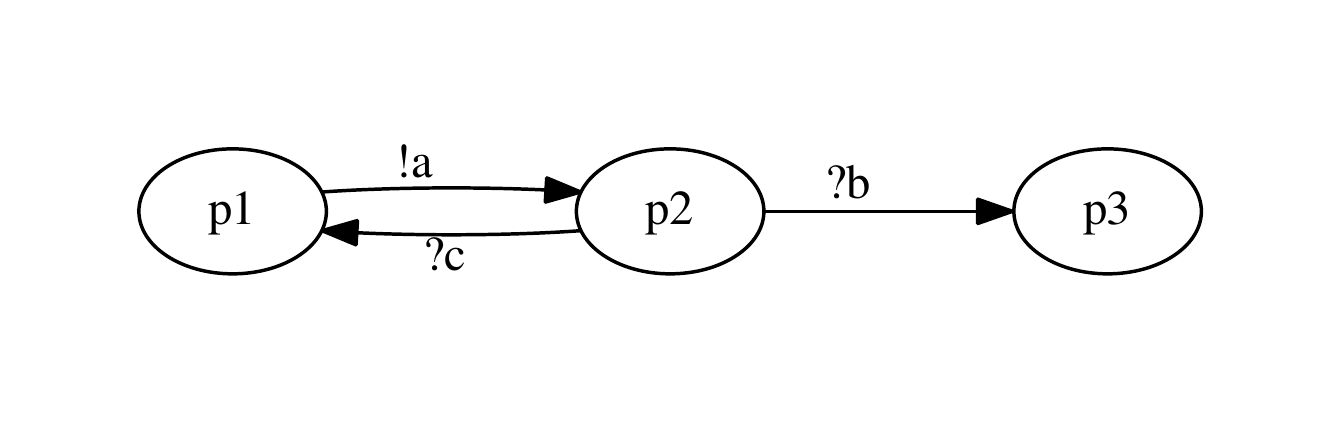}%
      \end{minipage}
    }%
  \end{minipage}
  \begin{minipage}{0.5\linewidth}
    \centering%
    \scalebox{0.95}{%
      \includegraphics[clip=true,trim=1.3cm 1.3cm 1.3cm 1.3cm,width=1.0\linewidth]{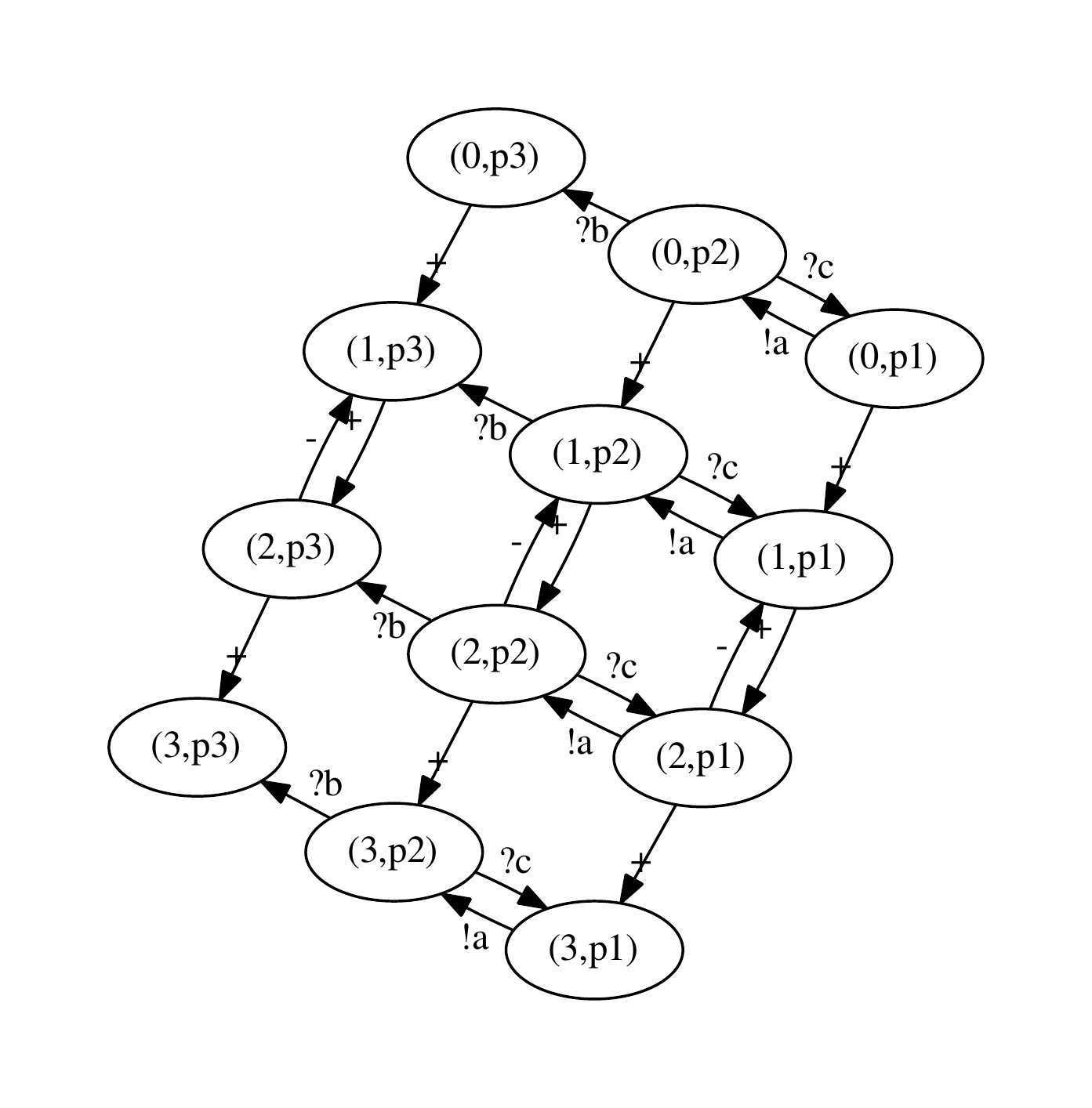}%
    }%
  \end{minipage}
  \caption{Output of %
    \code{l1.doDot} and \code{l2.toDot} (on the left), %
    and %
    \code{(l1\,|||\,l2).toDot} (on the right).}
  \label{fig:lts-par}%
\end{figure}%

In \RelEng, %
a finite LTS can be defined %
with the \code{LTS} constructor, %
by enumerating the state-(label-state) triples %
which compose its transition relation. %
For example:

{\small%
  \noindent%
  \begin{lrbox}{\mybgbox}%
    \begin{minipage}{\linewidth-2\fboxsep}%
      \begin{Verbatim}[commandchars=\\\{\},frame=single,framesep=0.5\fboxsep,framerule=0.8pt,rulecolor=\color{listingFrame}]
\PY{k}{val} \PY{n}{l1} \PY{k}{=} \PY{n+nc}{LTS}\PY{o}{(}\PY{n+nc}{List}\PY{o}{(}\PY{o}{(}\PY{l+m+mi}{0}\PY{o}{,} \PY{o}{(}\PY{l+s}{\PYZdq{}+\PYZdq{}}\PY{o}{,} \PY{l+m+mi}{1}\PY{o}{)}\PY{o}{)}\PY{o}{,} \PY{o}{(}\PY{l+m+mi}{1}\PY{o}{,} \PY{o}{(}\PY{l+s}{\PYZdq{}+\PYZdq{}}\PY{o}{,} \PY{l+m+mi}{2}\PY{o}{)}\PY{o}{)}\PY{o}{,} \PY{o}{(}\PY{l+m+mi}{2}\PY{o}{,} \PY{o}{(}\PY{l+s}{\PYZdq{}+\PYZdq{}}\PY{o}{,} \PY{l+m+mi}{3}\PY{o}{)}\PY{o}{)}\PY{o}{,} \PY{o}{(}\PY{l+m+mi}{2}\PY{o}{,} \PY{o}{(}\PY{l+s}{\PYZdq{}\PYZhy{}\PYZdq{}}\PY{o}{,} \PY{l+m+mi}{1}\PY{o}{)}\PY{o}{)}\PY{o}{)}\PY{o}{)}
\PY{k}{val} \PY{n}{l2} \PY{k}{=} \PY{n+nc}{LTS}\PY{o}{(}\PY{n+nc}{List}\PY{o}{(}\PY{o}{(}\PY{l+s}{\PYZdq{}p1\PYZdq{}}\PY{o}{,} \PY{o}{(}\PY{l+s}{\PYZdq{}!a\PYZdq{}}\PY{o}{,} \PY{l+s}{\PYZdq{}p2\PYZdq{}}\PY{o}{)}\PY{o}{)}\PY{o}{,} \PY{o}{(}\PY{l+s}{\PYZdq{}p2\PYZdq{}}\PY{o}{,} \PY{o}{(}\PY{l+s}{\PYZdq{}?b\PYZdq{}}\PY{o}{,} \PY{l+s}{\PYZdq{}p3\PYZdq{}}\PY{o}{)}\PY{o}{)}\PY{o}{,} \PY{o}{(}\PY{l+s}{\PYZdq{}p2\PYZdq{}}\PY{o}{,} \PY{o}{(}\PY{l+s}{\PYZdq{}?c\PYZdq{}}\PY{o}{,} \PY{l+s}{\PYZdq{}p1\PYZdq{}}\PY{o}{)}\PY{o}{)}\PY{o}{)}\PY{o}{)}
\end{Verbatim}
    \end{minipage}%
  \end{lrbox}%
  \colorbox{listingBG}{\usebox{\mybgbox}}%
}

\noindent%
The types of \code{l1} and \code{l2} are (respectively) %
\code{FiniteLTS[Int,String]} %
and %
\code{FiniteLTS[String,String]}: %
\ie, they are finite-state, finite-branching LTSs %
where states are \code{Int}egers (resp.~\code{String}s), and
labels are \code{String}s. %
The methods %
\code{l1.toDot} and \code{l2.toDot} return their graphs %
(shown on the left of \Cref{fig:lts-par}). %
The \code{|||} operator on LTSs %
returns the %
LTS whose states correspond to the parallel composition of its %
arguments' states, %
provided that the labels have the same type: %
\Cref{fig:lts-par} (on the right) shows the diagram of %
\code{(l1 ||| l2).toDot}. %
Such a composition is performed \emph{lazily}, %
thus avoiding (or delaying) state space explosion problems: %
the actual combinations of LTS states are generated only upon request.

A process can be simply retrieved from an LTS through one of its
states. %
For example: %

{%
  \noindent%
  \begin{lrbox}{\mybgbox}%
    \begin{minipage}{\linewidth-2\fboxsep}%
      \begin{Verbatim}[commandchars=\\\{\},frame=single,framesep=0.5\fboxsep,framerule=0.8pt,rulecolor=\color{listingFrame}]
\PY{k}{val} \PY{n}{p1} \PY{k}{=} \PY{n}{l2}\PY{o}{.}\PY{n}{process}\PY{o}{(}\PY{l+s}{\PYZdq{}p1\PYZdq{}}\PY{o}{)}
\end{Verbatim}
    \end{minipage}%
  \end{lrbox}%
  \colorbox{listingBG}{\usebox{\mybgbox}}%
}%

\noindent
In this case, we have that \code{p1} has type %
\code{FiniteProcess[String,String]} %
(\ie, a finite-state, finite-branching process %
where states are \code{String}s, and labels are \code{String}s as
well). %
As one might expect, %
\code{p1.state} has indeed value \codeStr{p1}. %
Moreover, %
\code{p1.lts} is \code{l2} %
--- \ie, the LTS inhabited by \code{p1}.

A process can be queried for its enabled transitions. %
In our example, \code{p1.transitions} has type %
\code{FiniteSet[String]}, %
and value \code{Set(\codeStr{!a})}. %
We can now let:

  \noindent%
  \begin{lrbox}{\mybgbox}%
    \begin{minipage}{\linewidth-2\fboxsep}%
      \begin{Verbatim}[commandchars=\\\{\},frame=single,framesep=0.5\fboxsep,framerule=0.8pt,rulecolor=\color{listingFrame}]
\PY{k}{val} \PY{n}{p1a} \PY{k}{=} \PY{n}{p1}\PY{o}{(}\PY{l+s}{\PYZdq{}!a\PYZdq{}}\PY{o}{)}\PY{o}{;}   \PY{k}{val} \PY{n}{p2} \PY{k}{=} \PY{n}{p1a}\PY{o}{.}\PY{n}{iterator}\PY{o}{.}\PY{n}{next}
\end{Verbatim}
    \end{minipage}%
  \end{lrbox}%
  \colorbox{listingBG}{\usebox{\mybgbox}}%

\noindent%
where \code{p1a} is the \code{FiniteSet} of continuations of
\code{p1} via transition \codeStr{!a}. %
In our example, %
\code{p1a} contains a single element, %
\ie the process corresponding to state \codeStr{p2} of \code{l2}: %
such a process is retrieved via \code{p1a}'s iterator%
\footnote{%
  Note that the same process can also be retrieved via %
  \code{l2.process(\codeStr{p2})}, %
  as we did for \code{p1} above.%
}, %
and assigned to \code{p2}. %
As expected, %
\code{p2.transitions} has value \code{Set(\codeStr{?b},\codeStr{?c})}.

Processes can be composed in parallel, %
similarly to LTSs (as shown above). %
Let:%

  \noindent%
  \begin{lrbox}{\mybgbox}%
    \begin{minipage}{\linewidth-2\fboxsep}%
      \begin{Verbatim}[commandchars=\\\{\},frame=single,framesep=0.5\fboxsep,framerule=0.8pt,rulecolor=\color{listingFrame}]
\PY{k}{val} \PY{n}{p01} \PY{k}{=} \PY{n}{l1}\PY{o}{.}\PY{n}{process}\PY{o}{(}\PY{l+m+mi}{0}\PY{o}{)} \PY{o}{|||} \PY{n}{p1}
\end{Verbatim}
    \end{minipage}%
  \end{lrbox}%
  \colorbox{listingBG}{\usebox{\mybgbox}}%

\noindent%
\code{p10} has type \code{FiniteProcess[(Int,String),String]} %
(\ie, each state is a \emph{pair} of \code{(Int,String)}, %
while labels remain \code{String}s). %
The transitions of \code{p01} are those of the LTS state \code{(0,p1)} %
in \Cref{fig:lts-par}: %
indeed, %
the same process could have been extracted with %
\code{(l1\,|||\,l2).process((0,\codeStr{p1}))}, %
and \code{p01.lts} is \code{l1\,|||\,l2}.

\subsection{CCS processes}
\label{sec:process-construct}

\RelEng implements \code{CCS}, %
which is the \emph{infinite} LTS %
whose states are \code{CCSTerm}s, %
labels are \code{CCSPrefix}es, %
and the (infinite)
transition relation corresponds to the CCS semantics.
Processes can be extracted from \code{CCS} as above, %
\ie with \code{CCS.process(s)} (where \code{s} is a \code{CCSTerm}), %
or letting \RelEng parse terms from strings:

{%
  \small%
  \noindent%
  \begin{lrbox}{\mybgbox}%
    \begin{minipage}{\linewidth-2\fboxsep}%
      \begin{Verbatim}[commandchars=\\\{\},frame=single,framesep=0.5\fboxsep,framerule=0.8pt,rulecolor=\color{listingFrame}]
\PY{k}{val} \PY{n}{ccs1} \PY{k}{=} \PY{n+nc}{CCS}\PY{o}{.}\PY{n}{process}\PY{o}{(}\PY{l+s}{\PYZdq{}rec(X)(!a.(?b + ?c.X))\PYZdq{}}\PY{o}{)} \PY{c+c1}{// Parses the CCSTerm from String}
\PY{k}{val} \PY{n}{ccs2} \PY{k}{=} \PY{n+nc}{CCS}\PY{o}{(}\PY{l+s}{\PYZdq{}?a.(t.!c.?a.!b + t.!b)\PYZdq{}}\PY{o}{)} \PY{c+c1}{// Shorthand. \PYZdq{}t\PYZdq{} is the internal action}
\end{Verbatim}
    \end{minipage}%
  \end{lrbox}%
  \colorbox{listingBG}{\usebox{\mybgbox}}%
}%

\noindent%
The type of \code{ccs1} and \code{ccs2} is %
\code{FiniteBranchingProcess[CCSTerm,CCSPrefix]} %
--- \ie, they are finite-branching %
(but \emph{not} necessarily finite-state) processes %
whose states are \code{CCSTerm}s, %
and whose transition labels are \code{CCSPrefix}es. %
Note that \code{ccs1} has, intuitively, %
the same transitions of process \code{p1} defined earlier: %
for example, %
\code{ccs1.transitions} is \code{Set($\ltsOut{a}$)}. %
There is, however, a difference: %
the CCS LTS distinguishes \code{CCSPrefix}es among %
\emph{input}, \emph{output} and \emph{internal} actions %
(respectively: $\ltsIn{a}, \ltsOut{a}, \ltsTau$), %
and this additional information %
(which is \emph{not} present for the simple string labels of process
\code{p1} above) %
is exploited by the \code{|||} operator %
to let two parallel CCS processes synchronise. %
For example, let:%

  \noindent%
  \begin{lrbox}{\mybgbox}%
    \begin{minipage}{\linewidth-2\fboxsep}%
      \begin{Verbatim}[commandchars=\\\{\},frame=single,framesep=0.5\fboxsep,framerule=0.8pt,rulecolor=\color{listingFrame}]
\PY{k}{val} \PY{n}{ccs12} \PY{k}{=} \PY{n}{ccs1} \PY{o}{|||} \PY{n}{ccs2}
\end{Verbatim}
    \end{minipage}%
  \end{lrbox}%
  \colorbox{listingBG}{\usebox{\mybgbox}}%

\noindent%
Here, \code{ccs12} has type %
\code{FiniteBranchingProcess[(CCSTerm,CCSTerm),CCSPrefix]}, %
and the value of %
\code{ccs12.transitions} is %
\code{Set($\ltsIn{a}$, $\ltsOut{a}$, $\ltsTau$)}. %
As expected, %
the $\ltsTau$-transition is generated by the synchronisation on
$\atom{a}$ %
--- and indeed, as shown in \Cref{fig:ccs12}, %
\code{ccs12($\ltsTau$)} returns%
\footnote{%
  Note that \code{ccs12($\ltsTau$)} and its return value %
  have been slightly edited for clarity, %
  and thus are \emph{not} valid Scala code.
}:%

  \noindent%
  \begin{lrbox}{\mybgbox}%
    \begin{minipage}{\linewidth-2\fboxsep}%
      \begin{Verbatim}[commandchars=\\\{\},frame=single,framesep=0.5\fboxsep,framerule=0.8pt,rulecolor=\color{listingFrame}]
\PY{n+nc}{Set}\PY{o}{(} \PY{o}{(} \PY{o}{(}\PY{o}{?}\PY{n}{b} \PY{o}{+} \PY{o}{(}\PY{o}{?}\PY{n}{c}\PY{o}{.}\PY{n}{rec}\PY{o}{(}\PY{n}{X}\PY{o}{)}\PY{o}{(}\PY{o}{!}\PY{n}{a}\PY{o}{.}\PY{o}{(}\PY{o}{?}\PY{n}{b} \PY{o}{+} \PY{o}{?}\PY{n}{c}\PY{o}{.}\PY{n}{X}\PY{o}{)}\PY{o}{)}\PY{o}{)}\PY{o}{)} \PY{o}{,} \PY{o}{(}\PY{n}{t}\PY{o}{.}\PY{o}{!}\PY{n}{c}\PY{o}{.}\PY{o}{?}\PY{n}{a}\PY{o}{.}\PY{o}{!}\PY{n}{b} \PY{o}{+} \PY{n}{t}\PY{o}{.}\PY{o}{!}\PY{n}{b}\PY{o}{)} \PY{o}{)} \PY{o}{)}
\end{Verbatim}
    \end{minipage}%
  \end{lrbox}%
  \colorbox{listingBG}{\usebox{\mybgbox}}%

\noindent%
The synchronisation mechanics are parametric at the LTS level %
--- and in particular, they are regulated by two methods:
\begin{itemize}
\item%
  \code{LTS.syncp(l1, l2)} is a predicate telling whether labels
  \code{l1} and \code{l2} can synchronise %
  (its default implementation is \code{false}, %
  thus only catering for interleaved executions, %
  as shown in \Cref{sec:from-lts-to-process}); %
\item%
  \code{LTS.syncLabel(l)} returns the new label emitted when
  synchronising on label \code{l} %
  (the default implementation is vacuous, %
  since \code{LTS.syncp()} is \code{false} by default).
\end{itemize}

Further details about the implementation of these methods in the case
of CCS are given in \Cref{sec:new-calculi}.

\subsection{From synchronous to asynchronous semantics}%
\label{sec:asynchrony}%

If \code{p} is an instance of \code{Process} %
(which is the main abstract class common to \emph{all} \RelEng
processes), %
then \code{p.async} is a new process obtained by pairing \code{p} with
an empty \emph{FIFO buffer}, %
represented as a \code{List}. %
\RelEng performs this transformation in a general, purely semantic
fashion%
\footnote{%
  Indeed, such an operation is performed at the LTS level: %
  if \code{l} is an \code{LTS}, %
  then \code{l.async} is the \code{LTS} with \code{l}'s states paired with a
  buffer; %
  if \code{s} is a state of \code{l}, %
  then \code{l.async.process((s, List()))} %
  is equal  to %
  \code{l.process(s).async}.  }: %
each \emph{output} label of \code{p} is appended to the buffer %
(with an internal transition), %
and the \emph{head} of the buffer enables a corresponding output
transition. %
This change is transparently reflected %
in the values returned by \code{p.async.transitions}. %
\iftoggle{techreport}{%
  If \code{p} has been created with the \code{CCS($\procP$)}
  constructor %
  (where $\procP$ is a CCS process), %
  then the semantics of \code{p.async} corresponds to
  $\procP\procQueue{}$, %
  as per \Cref{def:accs-semantics} %
  --- although there is no async-CCS-specific code for this
  functionality. %
}{}%
For example: 

  \noindent%
  \begin{lrbox}{\mybgbox}%
    \begin{minipage}{\linewidth-2\fboxsep}%
      \begin{Verbatim}[commandchars=\\\{\},frame=single,framesep=0.5\fboxsep,framerule=0.8pt,rulecolor=\color{listingFrame}]
\PY{k}{val} \PY{n}{ccs1a} \PY{k}{=} \PY{n}{ccs1}\PY{o}{.}\PY{n}{async}\PY{o}{;}   \PY{k}{val} \PY{n}{ccs2a} \PY{k}{=} \PY{n}{ccs2}\PY{o}{.}\PY{n}{async}
\end{Verbatim}
    \end{minipage}%
  \end{lrbox}%
  \colorbox{listingBG}{\usebox{\mybgbox}}%

\begin{figure}[t!]
  \centering%
  \scalebox{0.45}{%
    \includegraphics[clip=true,trim=1.3cm 1.3cm 1.3cm 1.3cm]{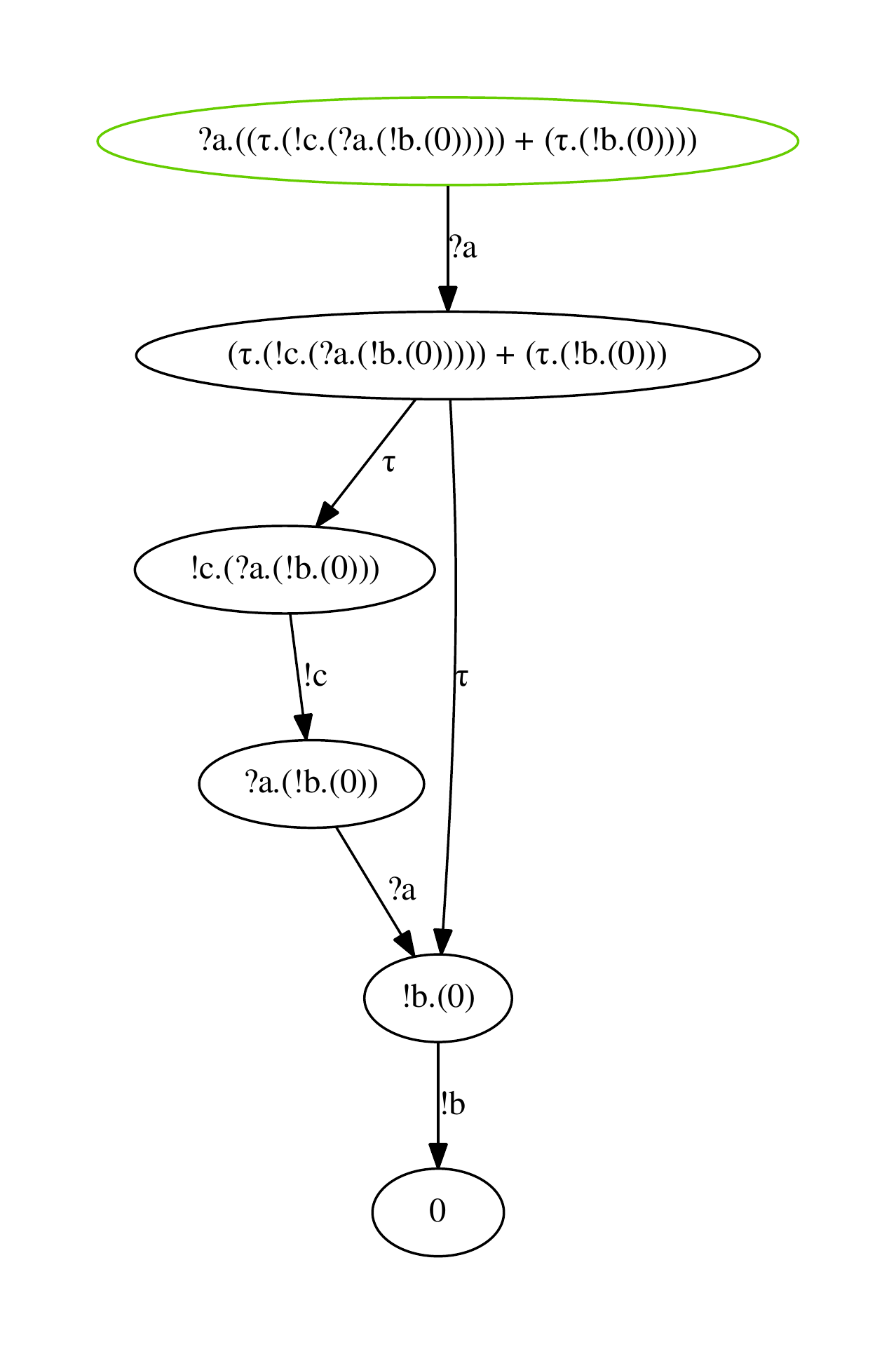}%
  }%
  \scalebox{0.45}{%
    \includegraphics[clip=true,trim=1.3cm 1.3cm 1.3cm 1.3cm]{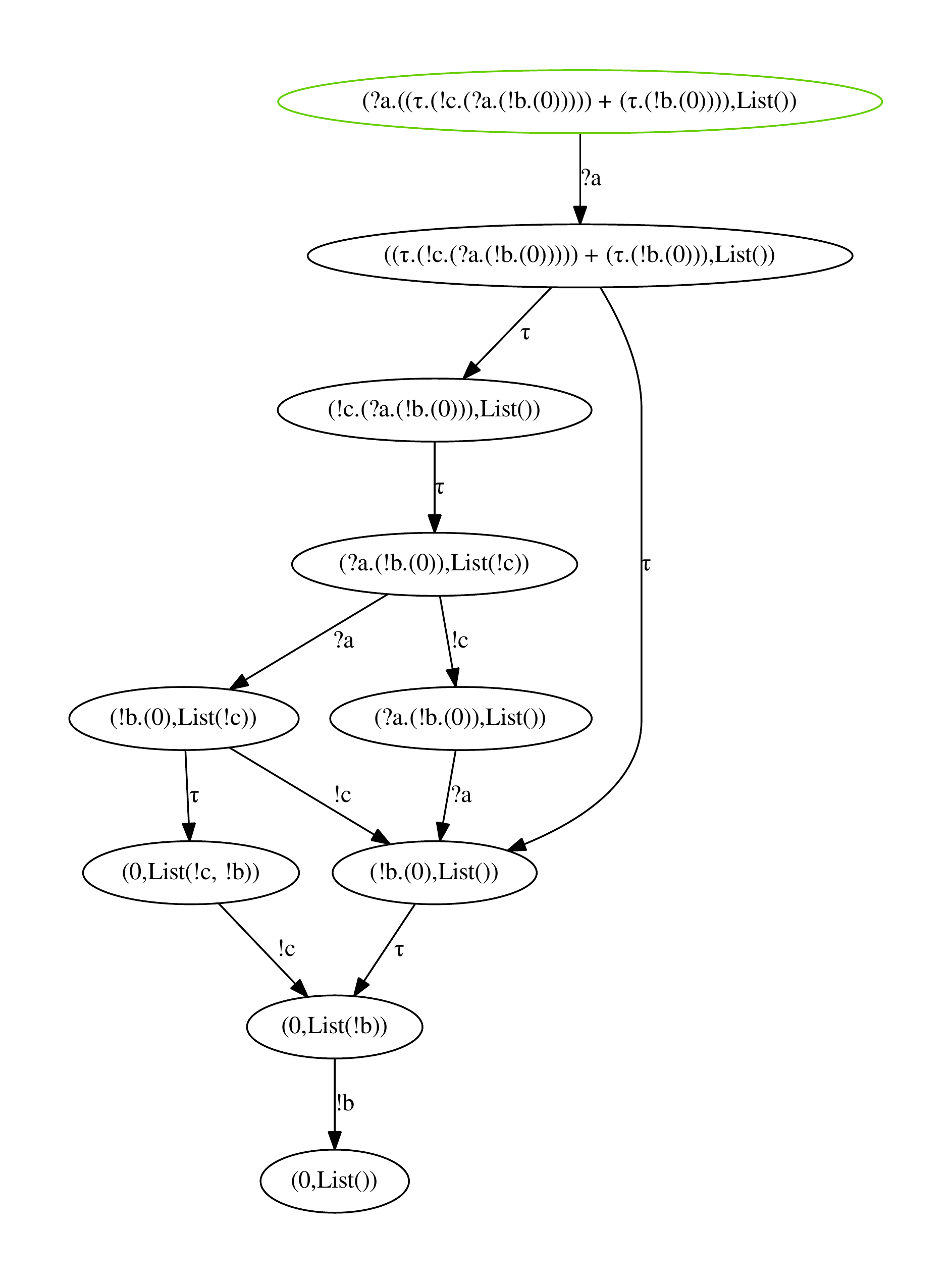}%
  }%
  \caption{Outputs of \code{ccs2.toDot()} (left) %
    and \code{ccs2.toDot()} (right).}
  \label{fig:ccs2-sync-async}%
  \vspace{-5mm}%
\end{figure}

\noindent%
\code{ccs1a} and \code{ccs2a} have type %
\code{FiniteBranchingProcess[(CCSTerm,Seq[CCSPrefix]),CCSPrefix]} %
(\ie, each state pairs a \code{CCSTerm} with a sequence of %
prefixes). %
The difference between \code{ccs2} and \code{ccs2a} is shown in
\Cref{fig:ccs2-sync-async}: %
%
%
it can be seen that, for example, the first $\ltsOut{c}$ transition
of \code{ccs2} becomes a $\ltsTau$ transition (with buffering) %
in \code{ccs2a}, %
and the head of the buffer is later consumed with a $\ltsOut{c}$ transition. %
Note, however, that there is %
an important difference between %
\code{ccs1} and \code{ccs1a}: %
while the former has a \emph{finite} number of states, %
the latter has \emph{infinite} states, %
due to the presence of recursion and unbounded buffers %
(the difference can be seen in \Cref{fig:ccs1-sync-async}). %
This is not a problem \emph{per se}, %
because, as remarked above, \RelEng ensures that process
transitions are expanded ``lazily''. 
%
Pairing a finite processes with an unbounded 
buffer %
reminds 
of Communicating Finite State Machines (CFSMs) \cite{BZ1983CFSM} %
--- and indeed, %
a CFSM-like interaction %
(modulo the different naming of labels) %
can be modeled with the composition %
\code{ccs1a\,|||\,ccs2a}, %
by filtering the states %
reachable via internal moves and synchronisations: %
the resulting \emph{finite} transition diagram is shown in
\Cref{fig:ccs1-ccs2-async-tau} %
(note that %
the ``unfiltered'' transition diagram of \code{ccs1a\,|||\,ccs2a} %
is \emph{in}finite).

\subsection{Adding new process calculi}
\label{sec:new-calculi}%

\RelEng has no ``hardwired'' notion of process calculus. %
A new process calculus %
with labelled semantics %
can be added to the framework in four steps: %
\emph{(i)} %
  define (or possibly reuse) a class \code{L} for its labels, %
\emph{(ii)} %
  define a class \code{T} for its terms, %
\emph{(iii)} %
  define a transition relation \code{R} %
  by deriving the class \code{Relation3[T,L,T]}, %
  and %
\emph{(iv)} %
  suitably derive the abstract class \code{LTS}, %
  using \code{T} and \code{L} respectively as state and label types %
  (specifying which labels are input/output/internal, %
  and how they synchronise), %
  and \code{R} as transition relation.
%
This very approach has been followed for implementing \code{CCS} under
\RelEng, %
as sketched below:

\begin{enumerate}[label=\enumroman]
\item%
  the base (abstract) class for CCS labels is \code{CCSPrefix}, %
  with one derived class for each concrete label type: %
  \code{CCSInPrefix}, \code{CCSOutPrefix}, and \code{CCSTau};
\item%
  the base (abstract) class for CCS terms is \code{CCSTerm}, %
  with one derivative for each syntactic production: %
  \code{CCSNil} (terminated process), %
  \code{CCSSeq} (prefix-guarded sequence), %
  \code{CCSPlus} (choice), \code{CCSPar} (parallel), %
  \code{CCSRec} (recursion), %
  \code{CCSVar} (recursion variable), %
  \code{CCSDel} (delimitation). %
  Such classes represent the CCS abstract syntax tree, %
  and they are instantiated by the CCS parser;
\item%
  the CCS semantics is implemented in the \code{CCSSemantics}
  singleton class. %
  Its core method is \code{apply(s:CCSTerm)}, %
  which returns the image of \code{s}, %
  \ie a binary \code{Relation[CCSPrefix,CCSTerm]} %
  containing the label-state transitions arising from \code{s}. %
  For example, %
  is \code{s} is a \code{CCSNil} instance, %
  the returned relation is empty; %
  if \code{s} is \code{CCSSeq(pfx:CCSPrefix, cont:CCSTerm)}, %
  the returned relation only contains the pair %
  \code{(pfx, cont)}, %
  and so on. %
  The other (more complex) cases exploit LTS-level or relational-level
  operators already provided by \RelEng\footnote{%
    The theory beyond such operators is sketched in %
    \Cref{sec:relational-calculus}.%
  }: %
  for example, %
  if \code{s} is \code{CCSPlus(term1, term2)}, %
  the return value is %
  \code{CCS.apply(term1) | CCS.apply(term2)}, %
  where \code{|} is the \emph{union} of the relations returned by
  invoking \code{apply()} on the two subterms: %
  as a consequence, %
  in the resulting relation, %
  a transition from \code{term1} leads to a
  continuation 
  which neglects \code{term2}, %
  and \emph{vice versa} %
  --- as expected by the standard behaviour of the CCS choice
  operator. %
  Instead, if \code{s} is \code{CCSPar(term1, term2)}, %
  the returned relation is created by directly reusing the
  syntax-independent, %
  LTS-level implementation of \code{|||} described in
  \Cref{sec:from-lts-to-process,sec:process-construct}%
  \footnote{%
    Such LTS operators are based on a \emph{relational} parallel composition
    operator: %
    the principle is the same sketched in %
    \Cref{sec:relational-calculus}.%
    \fxASwarning{More details? It seems too long to explain}%
  };
\item%
  finally, the CCS LTS is implemented in \code{CCS}, %
  which is a derivative of %
  \code{FiniteBranchingLTS[CCSTerm, CCSPrefix]}. %
  The \code{LTS.syncp(l1, l2)} method is overridden %
  so that it returns \code{true} whenever, for some string \code{a}, %
  \code{l1 == CCSInPrefix(a)} and \code{l2 == CCSOutPrefix(a)} %
  (or \emph{vice versa}); %
  moreover, the \code{LTS.syncLabel(l:CCSPrefix)} method is
  overridden %
  so that it returns \code{CCSTauPrefix()} %
  (\ie, each synchronisation causes the emission of a
  $\ltsTau$-prefix).
\end{enumerate}

With this approach, %
the CCS-specific code is mostly necessary for parsing terms, %
while the semantics of the operators is factored into several
syntax-independent classes; %
moreover, the implementation of \code{CCS.process()} %
and all the operations on CCS processes %
(\eg, \code{|||}, \code{.toDot()}, \code{.async},\ldots) %
are provided by the base abstract classes of \RelEng. %

We conclude this section noticing that, %
additionally to standard CCS syntactic constructs, %
\RelEng %
offers semantic operators allowing \eg process filtering %
(as we did for $\ltsTau$-reachable states in \Cref{sec:asynchrony}), %
and general sequencing: %
for all processes \code{p1}, \code{p2} with the same label type, %
\code{p1.seq(p2)} %
returns a process which behaves as \code{p1} until it
terminates, %
and then behaves as \code{p2}. %
These \emph{semantic} methods can be leveraged through the \RelEng
API, on \emph{all} LTSs and processes; %
if one wants to implement an additional process calculus with such
filtering/sequencing capabilities at the \emph{syntactic} level, %
then it is possible to simply reuse the underlying semantic
facilities, %
without reimplementing them.

Finally, we stress that, %
if two processes (notwithstanding their LTS) %
share the same label type, %
then they can synchronise, %
and their relations can be studied as shown in \Cref{sec:relations}.

\section{Behavioural relations}
\label{sec:relations}%

One of the goals of \RelEng is implementing and studying \emph{semantic}
relations, %
without syntactic limitations. 
\RelEng currently implements (bi)simulation, %
and some variants of 
\emph{progress} \cite{Castagna09toplas} %
and \emph{I/O compliance} \cite{BartolettiSZ14Concur}, %
\ie notions of ``correct'' interaction between processes. %
We exemplify the latter %
(the others are used similarly).

\begin{listing}[t!]
  \vspace{-0.3cm}%
  {%
    \footnotesize%
  \noindent%
  \begin{lrbox}{\mybgbox}%
    \begin{minipage}{\linewidth-2\fboxsep}%
      \begin{Verbatim}[commandchars=\\\{\},frame=single,framesep=0.5\fboxsep,framerule=0.8pt,rulecolor=\color{listingFrame}]
\PY{k}{val} \PY{n}{alice} \PY{k}{=} \PY{n+nc}{CCS}\PY{o}{(}\PY{l+s}{\PYZdq{}!aCoffee.?coffee.!pay + !aBeer.(?beer.!pay + ?no.!pay)\PYZdq{}}\PY{o}{)}
\PY{k}{val} \PY{n}{bartender} \PY{k}{=} \PY{n+nc}{CCS}\PY{o}{(}\PY{l+s}{\PYZdq{}rec(Y)(?aCoffee.!coffee.Y + ?aBeer.(!beer.Y + !no.Y) + ?pay)\PYZdq{}}\PY{o}{)}
\PY{k}{val} \PY{n}{ab} \PY{k}{=} \PY{n+nc}{IOCompliance}\PY{o}{.}\PY{n}{build}\PY{o}{(}\PY{n}{alice}\PY{o}{,} \PY{n}{bartender}\PY{o}{)}
\PY{k}{val} \PY{n}{aba} \PY{k}{=} \PY{n+nc}{IOCompliance}\PY{o}{.}\PY{n}{build}\PY{o}{(}\PY{n}{alice}\PY{o}{.}\PY{n}{async}\PY{o}{,} \PY{n}{bartender}\PY{o}{.}\PY{n}{async}\PY{o}{)}
\end{Verbatim}
    \end{minipage}%
  \end{lrbox}%
  \colorbox{listingBG}{\usebox{\mybgbox}}%
  }%
  \caption[A \RelEng example]{%
    \iftoggle{techreport}{%
      \RelEng example. %
      Alice and bartender CCS processes %
      are from \Cref{sec:running-example}.%
    }{%
      Alice and bartender example, %
      from \cite{BartolettiSZ14Concur}.%
    }%
  }%
  \label{lst:impl:releng:alice-bartender}%
\end{listing}

\subsection{Experiments with I/O compliance}%
\label{sec:compliance}%

\iftoggle{techreport}{}{%
  Intuitively, %
  two processes $\ltsSp,\ltsSq$ are I/O compliant %
  iff %
  the outputs of $\ltsSp$ are always matched by the inputs of $\ltsSq$ %
  (and \emph{vice versa}), %
  even after synchronisations and internal moves. %
}%
The \code{IOCompliance.build()} method takes two
\code{FiniteBranchingProcess} instances, %
and returns an \code{Either} object %
whose \code{Right} value is a \emph{finite} I/O compliance relation%
\iftoggle{techreport}{%
  (as per \Cref{def:compliance}.)
}{.} %
If $\ltsSp,\ltsSq$ are \emph{not} I/O compliant, %
the returned \code{Left} value is a \emph{counterexample}, %
\ie a pair of non-I/O compliant states.
%
Consider 
the first call to \code{IOCompliance.build()} in
Listing \ref{lst:impl:releng:alice-bartender}: %
since \code{alice} and \code{bartender} are I/O compliant, %
\code{ab}'s \code{Right} value is an I/O compliance relation %
containing the pair $(\text{\code{alice}}, \text{\code{bartender}})$; %
the same holds for \code{aba}, %
built on the \emph{asynchronous} versions of the two processes.

\begin{listing}[t!]
  {%
    \footnotesize%
  \noindent%
  \begin{lrbox}{\mybgbox}%
    \begin{minipage}{\linewidth-2\fboxsep}%
      \begin{Verbatim}[commandchars=\\\{\},frame=single,framesep=0.5\fboxsep,framerule=0.8pt,rulecolor=\color{listingFrame}]
\PY{k}{val} \PY{n}{aliceH} \PY{k}{=} \PY{n+nc}{CCS}\PY{o}{(}\PY{l+s}{\PYZdq{}!aCoffee.(?coffee | !pay)\PYZdq{}}\PY{o}{)}
\PY{k}{val} \PY{n}{bartenderL} \PY{k}{=} \PY{n+nc}{CCS}\PY{o}{(}\PY{l+s}{\PYZdq{}rec(Y)(?aCoffee.!coffee.Y + ?aBeer.(!beer.Y + !no.Y) + ?pay}
\PY{l+s}{                             + t . rec(Z)(?aCoffee.!coffee.Z + ?aBeer.!no.Z + ?pay))\PYZdq{}}\PY{o}{)}
\PY{k}{val} \PY{n}{aHbL} \PY{k}{=} \PY{n+nc}{IOCompliance}\PY{o}{.}\PY{n}{build}\PY{o}{(}\PY{n}{aliceH}\PY{o}{,} \PY{n}{bartenderL}\PY{o}{)}
\PY{k}{val} \PY{n}{aHbLa} \PY{k}{=} \PY{n+nc}{IOCompliance}\PY{o}{.}\PY{n}{build}\PY{o}{(}\PY{n}{aliceH}\PY{o}{.}\PY{n}{async}\PY{o}{,} \PY{n}{bartenderL}\PY{o}{.}\PY{n}{async}\PY{o}{)}
\end{Verbatim}
    \end{minipage}%
  \end{lrbox}%
  \colorbox{listingBG}{\usebox{\mybgbox}}%
  }%
  \caption[Another \RelEng example]{%
    \iftoggle{techreport}{%
      Another \RelEng example%
    }{%
      Another example from \cite{BartolettiSZ14Concur}%
    }: %
    Alice tries to grab the coffee and pay at the same time%
    \iftoggle{techreport}{%
      ; %
      the bartender, instead, may stop selling beer%
      (from \Cref{sec:running-example})%
    }{}.%
    \vspace{-0.4cm}%
  }%
  \label{lst:impl:releng:alice-late-bartender-nobeer}%
\end{listing}

Listing \ref{lst:impl:releng:alice-late-bartender-nobeer} shows more
examples%
\iftoggle{techreport}{%
  : %
  \code{aliceH} corresponds to $\procAliceCoffeeAsync$ in
  \Cref{sec:running-example}, %
  while \code{bartenderL} corresponds to $\procBartenderLateBeer$. %
}{%
  . %
}%
The \emph{second} call to \code{IOCompliance.build()} is successful
and returns \code{Right}, %
with an I/O compliance relation containing the \emph{asynchronous}
processes. %
\iftoggle{techreport}{%
  This result is coherent with
  \Cref{ex:typing-alice-bartender-safe-ii}. %
}{}%
The \emph{first} call to \code{IOCompliance.build()}, %
instead, %
is \emph{not} successful, %
and \code{aHbL} is the \code{Left} value below %
(edited for clarity): %

{%
  \small%
  \noindent%
  \begin{lrbox}{\mybgbox}%
    \begin{minipage}{\linewidth-2\fboxsep}%
      \begin{Verbatim}[commandchars=\\\{\},frame=single,framesep=0.5\fboxsep,framerule=0.8pt,rulecolor=\color{listingFrame}]
\PY{n+nc}{Left}\PY{o}{(} \PY{o}{(}\PY{o}{?}\PY{n}{coffee} \PY{o}{|} \PY{o}{!}\PY{n}{pay} \PY{o}{)}\PY{o}{,}
      \PY{o}{(}\PY{o}{!}\PY{n}{coffee}\PY{o}{.}\PY{n}{rec}\PY{o}{(}\PY{n}{Y}\PY{o}{)}\PY{o}{(}\PY{o}{?}\PY{n}{aCoffee}\PY{o}{.}\PY{o}{!}\PY{n}{coffee}\PY{o}{.}\PY{n}{Y} \PY{o}{+} \PY{o}{?}\PY{n}{aBeer}\PY{o}{.}\PY{o}{(}\PY{o}{!}\PY{n}{beer}\PY{o}{.}\PY{n}{Y} \PY{o}{+} \PY{o}{!}\PY{n}{no}\PY{o}{.}\PY{n}{Y}\PY{o}{)} \PY{o}{+} \PY{o}{?}\PY{n}{pay}
                      \PY{o}{+} \PY{n}{t}\PY{o}{.}\PY{n}{rec}\PY{o}{(}\PY{n}{Z}\PY{o}{)}\PY{o}{(}\PY{o}{?}\PY{n}{aCoffee}\PY{o}{.}\PY{o}{!}\PY{n}{coffee}\PY{o}{.}\PY{n}{Z} \PY{o}{+} \PY{o}{?}\PY{n}{aBeer}\PY{o}{.}\PY{o}{!}\PY{n}{no}\PY{o}{.}\PY{n}{Z} \PY{o}{+} \PY{o}{?}\PY{n}{pay}\PY{o}{)}\PY{o}{)}\PY{o}{)} \PY{o}{)}
\end{Verbatim}
    \end{minipage}%
  \end{lrbox}%
  \colorbox{listingBG}{\usebox{\mybgbox}}%
}%

\noindent%
The problem is that, %
after synchronising on $\atom{aCoffee}$,
\code{aliceH} and \code{bartenderL} reach the states %
inside \code{Left($\cdots$)}, %
where the $\ltsOut{pay}$ transition of the former %
is \emph{not} matched by a (weak) $\ltsIn{pay}$ 
of the latter. %
\iftoggle{techreport}{%
  This violates clause~\ref{item:compliance:i} of
  \Cref{def:compliance}).%
}{%
}

\subsection{Adding new compliance relations}%

Both \code{IOCompliance} and \code{Progress} %
\iftoggle{techreport}{(and their asymmetric versions) }{}%
are derivatives of an abstract, %
reusable class called \code{Compliance}. %
Intuitively, $\relR$ is a coinductive %
\emph{compliance relation} iff, %
whenever $(\ltsSp, \ltsSq) \in \mathord{\relR}$, then:
\begin{enumerate}[%
  noitemsep,topsep=0pt,parsep=0pt,partopsep=0pt,%
  label=\enumroman]%
\item%
  \label{item:compliance:i}%
  \code{pred($\ltsSp$,$\ltsSq$)} holds; %
  \quad%
  (where \code{pred} is given as a parameter) %
\item%
  \label{item:compliance:ii}%
  $\ltsSp \ltsMove{\ltsLab} \ltsSpi$ %
  and %
  $\ltsSq \ltsMove{\ltsLab'} \ltsSqi$ %
  and %
  $\ltsLab,\ltsLab'$ can synchronise %
  \;\;implies\;\; %
  $(\ltsSpi, \ltsSqi) \in \mathord{\relR}$; %
  \quad%
\item%
  \label{item:compliance:iii}%
  $\ltsSp \ltsWMove{} \ltsSpi$ %
  and %
  $\ltsSq \ltsWMove{} \ltsSqi$ %
  \;\;implies\;\; %
  $(\ltsSpi, \ltsSqi) \in \mathord{\relR}$. %
  \quad%
  (where $\ltsWMove{}$ represents 0 or more internal moves)
\end{enumerate}

\noindent %
\code{Compliance} implements the \code{.build()} method %
according to the definition above: %
given $(\ltsSp,\ltsSq)$, %
it ensures that a class-specific predicate
\code{pred} %
holds for $\ltsSp,\ltsSq$ (as per clause~\ref{item:compliance:i}), %
and then checks their reducts after synchronisation or internal moves %
(as per clauses \ref{item:compliance:ii} and \ref{item:compliance:iii}). %
\code{Compliance.build()} %
terminates %
when either no more states need to be checked, %
or \code{pred} is false: %
in the latter case, %
it returns a counterexample, %
as seen in %
\iftoggle{techreport}{%
  \Cref{sec:impl:releng:cex}%
}{%
  \Cref{sec:compliance}%
}. %
\iftoggle{techreport}{%
  For example, %
  the \code{IOCompliance}-specific predicate %
  matches clause~\ref{item:compliance:i} of \Cref{def:compliance}, %
  and \code{.build()} ensures that it holds for each pair of states in
  the relation. %
}{}%
\code{Progress}, \code{IOCompliance} and their variants %
are implemented by just changing \code{pred}, %
and new coinductive compliance relations can be added in the same way: %
\eg, %
the \emph{``Correct contract composition''} from \cite{Bravetti2007FSEN} %
(Def.~3) %
can be added by defining \code{pred($\ltsSp$,$\ltsSq$)} %
as %
\code{($\ltsSp$\,|||\,$\ltsSq$).wbarbs.contains($\checkmark$)} %
(where \code{.wbarbs} is the \code{Set} of weak barbs of a process, %
and $\checkmark$ is a label denoting success).

Note that \code{Compliance.build()} only implements a \emph{semi}-algorithm: %
hence, %
the method \emph{may} not terminate %
if one of the processes under analysis %
is infinite-state %
--- and in particular, %
if it can reduce, %
through internal moves, %
to an infinite number of distinct states. %
In such a situation, %
\RelEng may need to construct an \emph{infinite} compliance
relation, %
with an infinite search for states violating \code{pred}.
\iftoggle{techreport}{%
  \fxASwarning{lemma about $\compliant$ and $\ltsTau$ moves?}%
}{}%
Our Alice/bartender examples are infinite-state, %
but do \emph{not} generate infinite internal moves, %
and the semi-algorithm terminates. %
The risk of non-termination could be simply avoided %
by leveraging the types provided by \RelEng: %
for example, %
by only calling \code{Compliance.build()} on \code{FiniteProcess}
instances (\eg, through a simple wrapper). %
This would be a sufficient (but \emph{not} necessary) condition %
ensuring the termination of the method, %
albeit sacrificing cases such as the ones illustrated above. %
By letting \code{Compliance.build()} also accept
\code{FiniteBranchingProcess} arguments, %
\RelEng allows to experiment with behaviours for which the termination
of the method is not (yet) clear, %
or follows by some properties which are not easily captured by the
type system %
(\eg, the way inputs/outputs are interleaved in the Alice/bartender
example). %

\vspace{-0.4cm}%
\paragraph{Verifying relations.}%
\RelEng also implements the method \code{Compliance.check()}. %
Given an instance \code{r} of some \code{Compliance}-derived
relation, %
\code{r.check()} is \code{true} when each pair of states in \code{r} %
actually respects \code{pred} %
according to clause~\ref{item:compliance:i} above, %
and \code{r} contains all the pairs of states %
required by clauses~\ref{item:compliance:ii} and
\ref{item:compliance:iii}. %
Consider \eg %
Listing \ref{lst:impl:releng:alice-bartender}: %
\code{ab} is a \code{Right} value, %
and \code{ab.right.get.check()} is \code{true}%
\iftoggle{techreport}{%
  , %
  because for each pair of states, %
  clause~\ref{item:compliance:i} of \Cref{def:compliance} %
  is satisfied, %
  and the same holds for their $\ltsTau$-reducts %
  according to
  clauses~\ref{item:compliance:ii}--\ref{item:compliance:iv}%
}{%
}.  This also holds for \code{aba}, %
and \code{aHbLa} from
Listing \ref{lst:impl:releng:alice-late-bartender-nobeer}. %
It is important to note that \code{Compliance.build()} and
\code{Compliance.check()} are implemented \emph{separately}: %
the latter is intended as an independent verification method, %
also for relations which are defined ``by hand'' %
(\ie, directly as finite sets of pairs of states) %
\emph{without} resorting to their own \code{.build()}
method\footnote{%
  When debugging is enabled, %
  \RelEng runs \code{.check()} on \emph{each} relation created by
  \code{Compliance.build()}, %
  to test its code.%
}. %
For example, we can instantiate a \code{Progress} relation %
from an existing relation:

{%
  \small%
  \noindent%
  \begin{lrbox}{\mybgbox}%
    \begin{minipage}{\linewidth-2\fboxsep}%
      \begin{Verbatim}[commandchars=\\\{\},frame=single,framesep=0.5\fboxsep,framerule=0.8pt,rulecolor=\color{listingFrame}]
\PY{k}{val} \PY{n}{aHbLaProg} \PY{k}{=} \PY{n+nc}{Progress}\PY{o}{(}\PY{n}{aHbLa}\PY{o}{.}\PY{n}{right}\PY{o}{.}\PY{n}{get}\PY{o}{)} \PY{c+c1}{// Recall: aHbLa is an IOCompliance rel.}
\end{Verbatim}
    \end{minipage}%
  \end{lrbox}%
  \colorbox{listingBG}{\usebox{\mybgbox}}%
}%

\noindent%
and in this case \code{aHbLaProg.check()} holds %
--- \ie, notwithstanding its type, %
\code{aHbLa} is \emph{also} a progress relation.
Under this framework, %
if a new compliance relation is implemented as explained above %
(\ie, by deriving the \code{Compliance} class %
and providing a suitable class-specific \code{pred}), %
then %
synthesis (\code{.build()}) %
and verification (\code{.check()}) %
are obtained ``for free''. %
A similar framework is also in place for (bi)simulation.

\section{Conclusions and future work}
\label{sec:conclusion}%

In the current (early) stage of development, %
\RelEng offers a flexible and extensible platform %
allowing to define generic LTSs and processes, %
explore their (finite or infinite) state space %
and study their (bi)simulation and compliance relations. %
It offers general, syntax-independent operators for manipulating LTSs
and processes, %
on which specific process calculi can be implemented.

The most similar tool, %
albeit more CCS-centric, %
is \cite{Cleaveland1993CWB}, %
whose development stopped around 1999: %
hence, %
its obsolete dependencies and restrictive licensing terms %
make it very difficult to use and improve.
Another related tool is \emph{LTS Analyser} \cite{Magee2006} %
--- which is limited to finite-state processes; %
moreover, its development stopped around 2006, %
and its source code is not available.

It is possible to find some similarities between \RelEng and the
Process Algebra Compilers proposed in the '90s \cite{Cleaveland95Frontend}: %
\RelEng can be seen as a semantic backend on which a process calculus
can be ``compiled'' by suitably deriving some classes, %
and letting the parser instantiate them %
--- as sketched in \Cref{sec:new-calculi}. %
On the one hand, %
this approach makes the parser quite integrated into \RelEng, %
and not very suited for different backends; %
on the other hand, %
the tight integration allows to use parser combinators, %
thus obtaining easily maintainable, well-typed parsers.

Beyond representing and manipulating LTSs and processes, %
\RelEng also allows to explore them %
--- not unlike well-established model checking tools like mCRL2
\cite{Cranen13mCRLOverview} and CADP \cite{Garavel11CADP}. %
Besides being much smaller and less mature than such tools, %
\RelEng also has a different goal (being a \emph{framework} rather
than an application) %
and tries to keep a more \emph{semantic} foundation, %
in that it does not depend on (nor privileges) specific process languages. %
One intended usage scenario of \RelEng is the following: suppose you
want to introduce a new behavioural relation %
(say, I/O compliance), %
and you want to study it on some process algebra %
(say, asynchronous CCS), or on some processes whose specification is
provided directly as a set of state-label-state triples %
(e.g., from some industrial case study). %
One can achieve these goals by extending the \code{Compliance}
class, %
and applying it on LTSs and processes, as summarised in the paper. %
An alternative way would be that of %
\emph{(1)} %
  encoding asynchronous CCS or the given state-label-state triples
  into the process calculus and LTSs accepted by mCLR2 or CADP and
  their tools (proving that such an encoding is correct), %
  and %
\emph{(2)} %
  encode I/O compliance into \eg a $\mu$-calculus formula %
  (and, again, prove that such an encoding is correct).
%
Both alternatives are possible; however, we think that for the
scenario sketched above, the \RelEng framework allows users to obtain
quicker results.
We also believe that the relational calculus introduced in
\Cref{sec:relational-calculus} allows for greater flexibility and
reusability, %
\eg when carrying out experiments %
which require to %
combine LTSs and processes, %
or implement some process calculus.

Future work on \RelEng %
includes the addition of more relations, %
with a ``reusable'' approach to synthesis and verification %
similar to the one adopted for \code{Compliance} and (bi)simulation.
We also plan to fully formalise the relational calculus summarised in
\Cref{sec:relational-calculus}, and study its properties.
Moreover, we plan %
better support for multiparty interactions %
(currently provided via the \code{PCCS} calculus, %
not discussed here) %
and richer process calculi %
with time and value passing. %
We also plan to integrate \RelEng with Gephi \cite{Gephi}, %
thus providing a better user interface %
with interactive exploration of large transition diagrams.

{\small \paragraph{Acknowledgments.}
We would like to thank the anonymous reviewers for their detailed
comments and suggestions.
%
This work has been partially supported by: %
Aut.\ Reg.\ of Sardinia grants L.R.7/2007
CRP-17285 (TRICS) and P.I.A.\ 2010 (``Social Glue''),
by MIUR PRIN 2010-11 project ``Security Horizons'',
by EU COST Action IC1201
``Behavioural Types for Reliable Large-Scale Software Systems''
(BETTY), %
and by EPSRC grant EP/K011715/1.%
}

\bibliographystyle{eptcsini}
\bibliography{main}

\appendix%

\section{Figures}
\label{sec:figures}%

{%
  \centering%
  \includegraphics[clip=true,trim=1.3cm 1.3cm 1.3cm 1.3cm,width=0.9\linewidth]{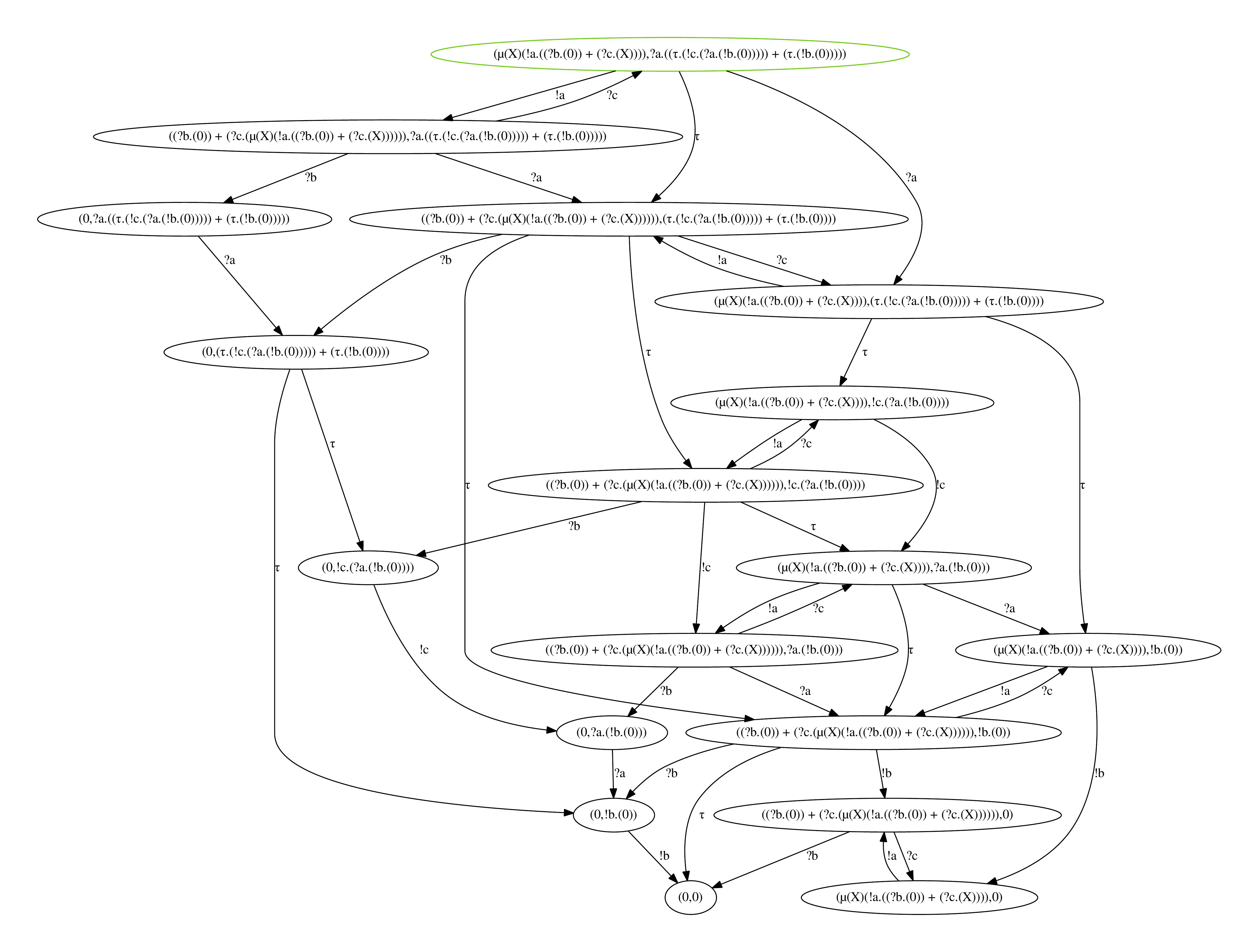}%
  \captionof{figure}{Output of \code{ccs12.toDot()}.}%
  \label{fig:ccs12}%
}%

\pagebreak

{%
  \centering%
  \includegraphics[clip=true,trim=1.3cm 1.3cm 1.3cm 1.3cm,width=0.7\linewidth]{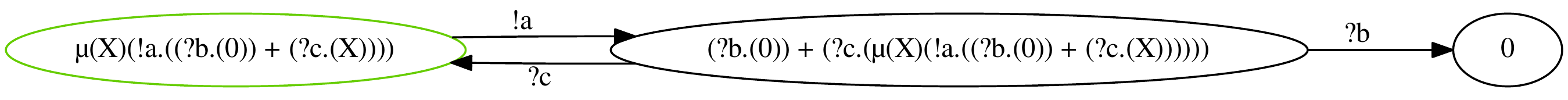}%
  \\[0.5cm]%
  \includegraphics[clip=true,trim=1.3cm 1.3cm 1.3cm 1.3cm,width=0.8\linewidth]{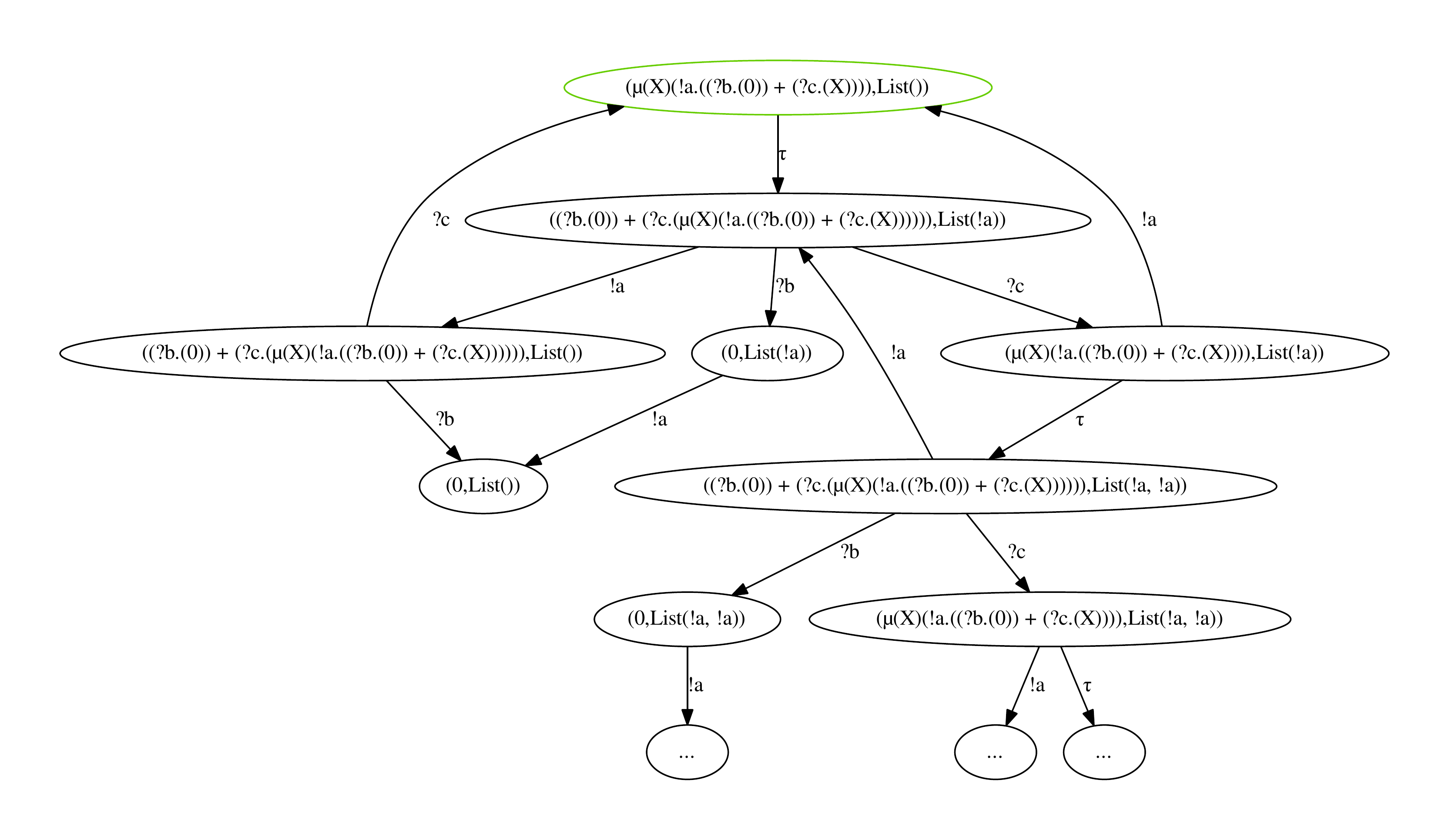}%
  \captionof{figure}{Output of \code{ccs1.toDot()} (top) %
    and \code{ccs1a.toDot(maxDepth=Finite(4))} (bottom).}%
  \label{fig:ccs1-sync-async}%
}%

\bigskip%

{%
  \centering%
  \includegraphics[clip=true,trim=1.3cm 1cm 1.3cm 1cm,width=1.0\linewidth]{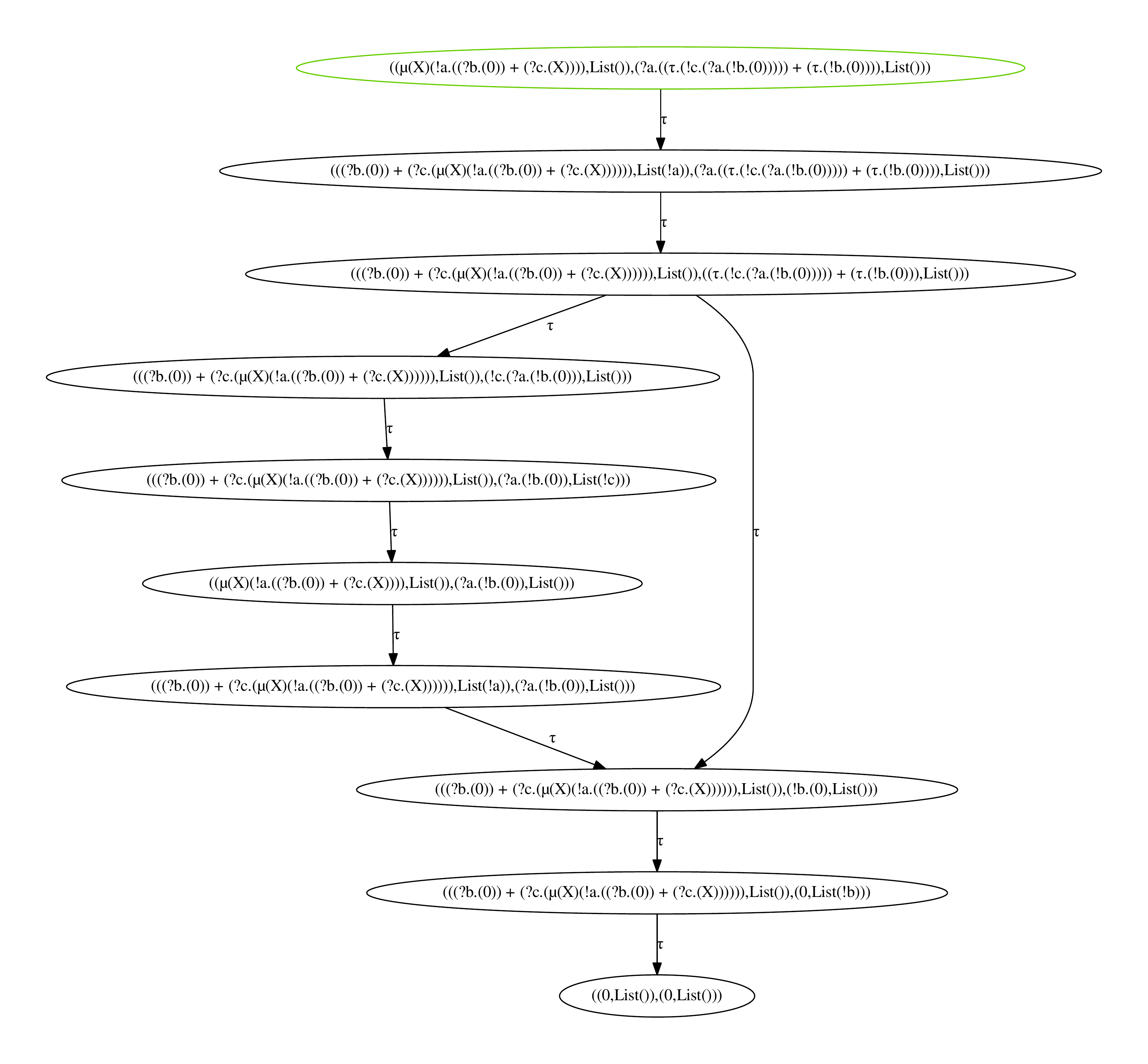}%
  \captionof{figure}{Output of %
    \code{(ccs1a ||| ccs2a).filter(l => l.isTau).toDot()}. %
    Note that $\ltsTau$-transitions generated by synchronisations %
    cause the reduction of buffers %
    --- \ie, the output at the head of a buffer is consumed by an
    input of the other process.%
  }%
  \label{fig:ccs1-ccs2-async-tau}%
}%
%

\ifdraft{%
  \pagebreak%
  \listoffixmes%
}{}

\end{document}